\documentclass[prd,aps,showkeys,superscriptaddress,two column]{revtex4-1}
\usepackage{amsmath}
\usepackage{amssymb}
\usepackage{amsthm}
\usepackage{mathrsfs}
\usepackage{graphicx}
\usepackage{fancyhdr}
\usepackage{array}
\usepackage{simplewick}
\usepackage{latexsym}
\usepackage[all]{xy}
\usepackage{eufrak}
\usepackage{euscript}
\usepackage{enumerate}
\usepackage{dsfont}
\usepackage{slashed}
\usepackage{hyperref}

\begin{document}

%%%%%%%%%%%%%%%%%%%%%%%%%%%%%%%%%%%%%%%%%%%%%%%%%%%%%%%%%%%%%

\title{Massive photons and Dirac monopoles: electric condensate and magnetic confinement}

\author{M. S. Guimaraes}
\email{msguimaraes@uerj.br}
\affiliation{Instituto de F\'\i sica, Universidade do Estado do Rio de Janeiro, 20550-013, Rio de Janeiro, Brazil}

\author{R. Rougemont}
\email{romulo@if.ufrj.br}
\affiliation{Instituto de F\'\i sica, Universidade Federal do Rio de Janeiro, 21941-972, Rio de Janeiro, Brazil}

\author{C. Wotzasek}
\email{clovis@if.ufrj.br}
\affiliation{Instituto de F\'\i sica, Universidade Federal do Rio de Janeiro, 21941-972, Rio de Janeiro, Brazil}

\author{C. A. D. Zarro}
\email{carlos.zarro@if.ufrj.br}
\affiliation{Instituto de F\'\i sica, Universidade Federal do Rio de Janeiro, 21941-972, Rio de Janeiro, Brazil}

%\date{\today}

\begin{abstract}
We use the generalized Julia-Toulouse approach (GJTA) for condensation of topological currents (charges or defects) to argue that massive photons can coexist consistently with Dirac monopoles. The Proca theory is obtained here via GJTA as a low energy effective theory describing an electric condensate and the mass of the vector boson is responsible for generating a Meissner effect which confines the magnetic defects in monopole-antimonopole pairs connected by physical open magnetic vortices described by Dirac brane invariants, instead of Dirac strings.
\end{abstract}

%\pacs{Valid PACS appear here}
% PACS, the Physics and Astronomy Classification Scheme.
% Valid PACS numbers may be entered using the \verb+\pacs{#1} command.

\keywords{Monopoles, massive electrodynamics.}
% Use showkeys class option if keyword display desired

%%%%%%%%%%%%%%%%%%%%%%%%%%%%%%%%%%%%%%%%%%%%%%%%%%%%%%%%%%%%%

\maketitle

\section{Introduction}
\label{sec:introduction}

In his seminal work \cite{dirac}, Dirac established a theory of magnetic monopoles interacting with massless vector bosons, from which emerged a possible explanation for the electric charge quantization observed in Nature: the mere existence of a monopole would imply in the quantization of the electric charge in multiples of the inverse of the magnetic charge, what is based on the consistency condition for the magnetic Dirac string to be unobservable at the quantum level. Since then, the physics involving Dirac monopoles has been proved to be useful also to investigate other physical scenarios \cite{mvf,jt-cho,ripka}.

Our aim in this work is to generalize the Dirac's non-minimal prescription for the case where the vector bosons are massive, with the hope to clarify some misunderstandings found in the literature, like the claims that Dirac monopoles and massive photons cannot coexist and that the Dirac strings would become observable when the vector bosons are massive \cite{joshi}.

One of the main points involved in this issue regards the fact that the Dirac theory of monopoles was developed in the context of massless vector bosons and its extension to the case of massive photons is not immediate. Another key point refers to the very general observation that a massive photon generates a Meissner effect, which confines magnetic probe sources. Together with these observations, one must also keep in mind that, since the Dirac strings are unphysical artifacts used to introduce monopoles in a theory with a single gauge potential defined over the whole spacetime, except at the location of the world-surfaces of these strings, there are no physical processes that could turn these Dirac branes into observables: this point is in fact a consistency condition that must be always satisfied in order to keep the consistency of the formalism. These basic observations can be gathered together through the use of a generalization of the so-called Julia-Toulouse approach for condensation of topological currents (charges or defects).

The original Julia-Toulouse approach \cite{jt,qt} is a prescription used to construct a low energy effective theory for a system with condensed charges or defects, having previous knowledge of the model describing the system in the regime where these sources are dilutely distributed through the space and also of the symmetries expected for the regime where the charges or the defects condense. Based mainly on \cite{jt,qt}, and taking also into account the ideas developed in \cite{banks,mvf} regarding the formulation of ensembles of charges and defects, we introduced in \cite{artigao,mbf-m} a generalization of the Julia-Toulouse approach, whose main feature is a careful treatment of a local symmetry which we call as the Dirac brane symmetry, which is independent of the usual gauge symmetry \cite{mvf}, and consists in the freedom of deforming the Dirac strings without any observable consequences. In what follows, we are going to call this generalized prescription as the generalized Julia-Toulouse approach (GJTA).

In the present work we shall follow a very general strategy to obtain a consistent formulation of the Proca theory in the presence of external monopoles. We begin with the Maxwell theory in the presence of diluted electric charges and introduce external magnetic defects through the original Dirac's non-minimal substitution, which can be safely applied to massless gauge theories. We then use the GJTA to construct the Proca theory in the regime where the electric charges condense, getting the correct definition of the massive electrodynamics in the presence of Dirac monopoles. Through this process, we shall see that due to the Dirac's veto \cite{dirac}, the Dirac branes are effectively removed from the formalism in the electric condensed regime, giving place to physical open magnetic vortices with a monopole-antimonopole pair in their ends. These open vortices are described by Dirac brane invariants corresponding to the confining magnetic flux tubes. In particular, since the magnetic probe sources are confined in this scenario due to the Meissner effect associated to the mass acquired by the vector boson as a result of the electric condensation process, it is impossible to introduce isolated magnetic defects into the massive electrodynamics, the only possibility being the introduction of mesonic monopole-antimonopole pairs (as far as we know, this conclusion was firstly explicitly pointed out in \cite{ahrens}). However, contrary to the usual claim found in the literature \cite{qt,mvf,joshi,ahrens,ripka,antonov}, the monopoles with opposite magnetic charges in these pairs are not connected by Dirac strings, but instead, they are connected by physical confining magnetic flux tubes described by Dirac brane invariants and we show how these structures emerge in the formalism by taking the Dirac brane symmetry carefully into account.

\section{Massive electrodynamics and Dirac monopoles}
\label{sec:proca}

We are going to work in $(3+1)$-dimensional Minkowski spacetime $\mathbb{R}^{1,3}$ and make use of natural units with $c=\hbar=1$.

The partition function of the Maxwell theory in the presence of diluted eletric charges and magnetic monopoles is given by:
\begin{align}
Z_d[J_1,j_1]&=\int_{G.F.}\mathcal{D}A_1 \exp\left\{i\int_{\mathbb{R}^{1,3}}\left[-\frac{1}{2}(dA_1 -g*\chi_2)\right.\right.\nonumber\\
&\left.\left.\wedge*(dA_1-g*\chi_2) - eA_1\wedge *J_1\right]\right\},
\label{eq:1}
\end{align}
where $J_1=\delta\Sigma_2$ is the topological electric current which localizes the world-line of the electric charge $e$, the physical boundary of the world-surface of the electric Dirac string localized by the Chern-Kernel $\Sigma_2$ and $j_1=\delta\chi_2$ is the topological magnetic current which localizes the world-line of the magnetic charge $g$, the physical boundary of the world-surface of the magnetic Dirac string localized by the Chern-Kernel $\chi_2$. The acronym ``G.F.'' stands for some ``gauge fixing'' procedure that must be used at some stage of the calculations.

As discussed in \cite{mbf-m}, the magnetic Dirac brane symmetry corresponds to the local invariance of (\ref{eq:1}) under deformations of the magnetic Dirac branes that keep fixed their physical boundaries corresponding to the monopole currents and also satisfies the Dirac's veto \cite{dirac,felsager}, which prohibits the magnetic Dirac branes of crossing the electric world-lines. This local symmetry implies in the Dirac charge quantization condition \cite{dirac,mvf}, $eg=2\pi n$, $n\in\mathbb{Z}$, as a consistency condition for the invisibility of the Dirac branes, which are unphysical.

Let us work with the electromagnetic dual of (\ref{eq:1}). For this sake, we make use of the master representation of (\ref{eq:1}):
\begin{align}
Z_d[J_1,j_1]&=\int_{G.F.}\mathcal{D}A_1\mathcal{D}G_2\exp\left\{i\int_{\mathbb{R}^{1,3}}\left[\frac{1}{2} 
G_2\wedge*G_2+\right.\right.\nonumber\\
&\left.\left.-G_2\wedge*(dA_1-g*\chi_2)-e A_1\wedge *J_1\right]\right\},
\label{eq:2}
\end{align}
from which we can return to the original representation (\ref{eq:1}) after integrating out the auxiliary field $G_2$. Instead of this, we integrate out the gauge field $A_1$ in (\ref{eq:2}), obtaining the dual representation:
\begin{align}
Z_d[J_1,j_1]&=\int\mathcal{D}G_2\,\delta\left[d*G_2+e*J_1\right]\nonumber\\
&\exp\left\{i\int_{\mathbb{R}^{1,3}}\left[\frac{1}{2} 
G_2\wedge*G_2-gG_2\wedge\chi_2\right]\right\}\nonumber\\
&=\int_{G.F.}\mathcal{D}C_1 \exp\left\{i\int_{\mathbb{R}^{1,3}}\left[-\frac{1}{2}(dC_1-e*\Sigma_2)\right.\right.\nonumber\\
&\left.\left.\wedge*(dC_1-e*\Sigma_2) + gC_1\wedge *j_1-eg*\Sigma_2\wedge*\chi_2\right]\right\}\nonumber\\
&=\int_{G.F.}\mathcal{D}C_1 \exp\left\{i\int_{\mathbb{R}^{1,3}}\left[-\frac{1}{2}(dC_1-e*\Sigma_2)\right.\right.\nonumber\\
&\left.\left.\wedge*(dC_1-e*\Sigma_2) + gC_1\wedge *j_1\right]\right\},
\label{eq:3}
\end{align}
where the dual gauge field $C_1$ has emerged by solving the functional constraint $d*G_2=-e*J_1\Rightarrow *G_2=dC_1-e*\Sigma_2$ and, in passing to the last line of (\ref{eq:3}), we used that $-eg\int_{\mathbb{R}^{1,3}}*\Sigma_2\wedge*\chi_2=-egN$, where $N$ is an integer corresponding to the intersection number between the electric and magnetic Dirac branes, such that, due to the Dirac charge quantization condition, the complex exponential of this term gives 1 and makes no contribution in the partition function \cite{mvf,dafdc}. The dual representation (\ref{eq:3}) is physically equivalent to the original representation (\ref{eq:1}), but here the couplings are inverted: the dual gauge field couples minimally to the monopole currents and non-minimally to the electric charges. Hence, from the point of view of the dual gauge field, the electric Dirac branes are seen as defects, being $C_1$ and $dC_1$ singular over these branes. Notice, however, that the non-minimal coupling $(dC_1-e*\Sigma_2)$, which represents the physical electromagnetic fields, is regular everywhere, since the singularity of $dC_1$ is exactly canceled out by the singular term $*\Sigma_2$ \cite{mvf,ripka}.

At this point, we are ready to apply the GJTA and consider the effects of a electric charge condensation in this system. The condensation of electric charges is represented here by a proliferation of the electric world-lines, which implies in a proliferation of the electric Dirac branes from which these world-lines are boundaries. Due to the proliferation of the electric Dirac branes, the dual gauge field becomes ill-defined in almost the whole space and its degrees of freedom are not adequate to describe the system in the electric condensed regime. However, the non-minimal coupling remains regular everywhere. The GJTA in this picture consists in taking the regular non-minimal coupling as a new field describing the low energy excitations of the electric condensate \cite{qt}:
\begin{align}
(dC_1-e*\Sigma_2)\stackrel{\textrm{cond.}}{\longmapsto}m H_2,
\label{eq:4}
\end{align}
where $m$ is a phenomenological mass scale associated to the electric condensate. Notice that the prescription (\ref{eq:4}) effectively promotes a dynamical term for the massless 1-form gauge field $C_1$ describing the system in the diluted regime to a mass term for the 2-form Kalb-Ramond field $H_2$ describing the system in the condensed regime: this \emph{rank-jumping} of the field describing the excitations of the theory and the associated \emph{mass gap generation} constitute a signature of the \emph{condensation of topological currents} in the picture where the condensing currents are non-minimally coupled to the gauge field describing the theory in the diluted regime \cite{jt,qt,dafdc,artigao}.

For a consistent implementation of the prescription (\ref{eq:4}) into (\ref{eq:3}), in order to obtain the partition function for the electric condensed regime in the dual picture, we must apply (\ref{eq:4}) also to the minimal coupling appearing in (\ref{eq:3}). For this sake, we must reveal the structure of the non-minimal coupling inside the minimal coupling term:
\begin{align}
g\int_{\mathbb{R}^{1,3}}C_1\wedge*j_1&=g\int_{\mathbb{R}^{1,3}}C_1\wedge**d(*\chi_2-d*\lambda_3)\nonumber\\
&=g\int_{\mathbb{R}^{1,3}}dC_1\wedge(*\chi_2-d*\lambda_3)\nonumber\\
&\doteq g\int_{\mathbb{R}^{1,3}}(dC_1-e*\Sigma_2)\wedge(*\chi_2-d*\lambda_3),
\label{eq:5}
\end{align}
where in the first line we made explicit the Dirac brane ambiguity involved in the definition of the monopole currents and in the passage to the last line we added two intersection numbers that do not contribute in the Boltzmann factor due to the Dirac charge quantization condition (hence, this last equality only holds inside the partition function). Applying the prescription (\ref{eq:4}) into (\ref{eq:5}), we get:
\begin{align}
g\int_{\mathbb{R}^{1,3}}C_1\wedge*j_1\stackrel{\textrm{cond.}}{\longmapsto}mg
\int_{\mathbb{R}^{1,3}} H_2\wedge*L_2,
\label{eq:6}
\end{align}
where we defined the magnetic Dirac brane invariant:
\begin{align}
*L_2 := *\chi_2 - *\Omega_2 = *\chi_2 - d*\lambda_3,
\label{eq:7}
\end{align}
where $\Omega_2=\delta\lambda_3$ is a topological current, which is identically conserved due to the nilpotency of the codifferential, $\delta\Omega_2=0$: it describes a current density of closed magnetic vortices associated with regions of the space where the electric condensate has not been established, as will become clear in the discussion following eq. (\ref{eq:10}). Notice also from (\ref{eq:5}) that without the Dirac charge quantization condition it would be impossible to obtain (\ref{eq:6}) and (\ref{eq:7}) in the electric condensed regime: the Dirac charge quantization condition is, therefore, a necessary condition for a consistent formulation of the system in the electric condensed regime via GJTA, being intrinsically related to the establishment of the magnetic Dirac brane invariants in this phase. When we deform the magnetic Dirac strings, $*\chi_2\mapsto *\chi_2+d*\tau_3$, where $\tau_3$ is the volume spanned in $\mathbb{R}^{1,3}$ by the deformation of the world-surface of the Dirac string, keeping fixed its boundary, the $\lambda_3$-term transforms as $*\lambda_3\mapsto *\lambda_3 + *\tau_3$, such that $*L_2$ is kept invariant under Dirac string deformations. This is the reason why we call it a Dirac brane invariant (distribution). Also the $H_2$ field describing the electric condensate is a Dirac brane invariant (field), since it is defined in (\ref{eq:4}) in terms of a non-minimal coupling, which is also an invariant \cite{mvf,artigao}.

To complete the construction of the effective field theory describing the lowest-lying modes of the electric condensate, we take a derivative expansion of the electric condensate field, $H_2$, and retain only the dominant contribution at low energies, which is the term with lowest order in derivatives satisfying the relevant symmetries of the system (in this case, Lorentz and $C$, $P$ and $T$). In this way, we obtain for the electric condensed regime the following low energy effective theory in the dual picture:
\begin{align}
Z_c[j_1]&=\sum_{\left\{*\lambda_3\right\}}\int\mathcal{D}H_2 \exp\left\{i\int_{\mathbb{R}^{1,3}}\left[\frac{1}{2}
dH_2\wedge*dH_2 +\right.\right.\nonumber\\
&\left.\left. -\frac{m^2}{2}H_2\wedge*H_2+mgH_2\wedge*L_2\right]\right\},
\label{eq:8}
\end{align}
where the ensemble of internal defects $\left\{*\lambda_3\right\}$ represents the contribution of the magnetic vortices in the system: this constitutes a generalization of the effective theory obtained in \cite{qt,antonov}, where the magnetic vortex contribution is missing. The electromagnetic dual of (\ref{eq:8}) is obtained by making use of the master representation (see also \cite{ripka,antonov}):
\begin{align}
Z_c[j_1]&=\sum_{\left\{*\lambda_3\right\}}\int\mathcal{D}H_2\mathcal{D}(*G_3) \exp\left\{i\int_{\mathbb{R}^{1,3}}\left[-\frac{1}{2}G_3\wedge*G_3+\right.\right.\nonumber\\
&\left.\left.+G_3\wedge*dH_2-\frac{m^2}{2}H_2\wedge*H_2+mgH_2\wedge*L_2\right]\right\},
\label{eq:9}
\end{align}
from which we can return to (\ref{eq:8}) by integrating out the auxiliary field $G_3$. Instead of this, we integrate out the Kalb-Ramond field $H_2$, obtaining the dual representation of the partition function (\ref{eq:8}) describing the electric condensed regime:
\begin{align}
&Z_c[j_1]=\sum_{\left\{*\lambda_3\right\}}\int\mathcal{D}(*G_3) \exp\left\{i\int_{\mathbb{R}^{1,3}}\left[-\frac{1}{2m^2}(d(*G_3)+\right.\right.\nonumber\\
&\left.\left.-mg*L_2)\wedge*(d(*G_3)-mg*L_2)+\frac{1}{2}(*G_3)\wedge*(*G_3)\right]\right\}\nonumber\\
&\;\;\;\;\;\;\;\;\;=\sum_{\left\{*\lambda_3\right\}}\int\mathcal{D}A_1 \exp\left\{i\int_{\mathbb{R}^{1,3}}\left[-\frac{1}{2}(dA_1-g*L_2)\right.\right.\nonumber\\
&\;\;\;\;\;\;\;\;\;\left.\left.\wedge*(dA_1-g*L_2)+\frac{m^2}{2}A_1\wedge *A_1\right]\right\},
\label{eq:10}
\end{align}
where in the passage to the last line we defined $*G_3=:mA_1$ (notice that the $A_1$ field here is a massive vector excitation of the electric condensate, and not the original massless gauge potential of eq. (\ref{eq:1})). Equation (\ref{eq:10}) defines the partition function of the Proca theory in the presence of external probe magnetic monopoles in a consistent way, as we are going to discuss now.

\begin{figure}[h]  % label dentro da caption para assegurar referência correta no corpo do texto
       \centering
       \includegraphics[scale=0.5]{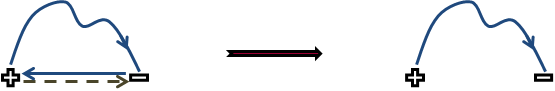}
       \caption{The magnetic flux inside the Dirac string (gray dashed arrow) is $g$, while the magnetic flux inside the closed vortex (blue solid line) is $2\pi n/e$. The Dirac's veto imposes that the only place allowed for the magnetic Dirac string in the electric condensate is the interior of a closed magnetic vortex formally connected to a monopole-antimonopole pair and, therefore, the Dirac charge quantization condition removes the Dirac string from the formalism together with the unphysical part of this vortex connected to the monopole-antimonopole pair, leaving as the result a physical open magnetic vortex corresponding to a confining flux tube for the monopole-antimonopole probe configuration. \label{fig:1}}
\end{figure}

Let us first address the physical interpretation of the term $\Omega_2=\delta\lambda_3$ featured in the definition of the magnetic Dirac brane invariant (\ref{eq:7}), as promised earlier. For this sake, we begin by setting $\chi_2=0$ in (\ref{eq:7}), thus considering the Proca theory, which is the London limit of the Abelian Higgs model describing a relativistic superconductor \cite{mvf,ripka}, in the absence of monopoles. Then, we have for the electromagnetic fields in the kinetic term in (\ref{eq:10}), a closed magnetic flux, $g*L_2=\frac{2\pi n}{e}d*\lambda_3$, which is quantized in integer multiples of $\frac{2\pi}{e}$: this is just the well known contribution of the magnetic vortices inside a type-II superconductor. Hence, as stated before, $\Omega_2$ is a topological current density of closed magnetic vortices and $\lambda_3$ is the associated Chern-Kernel. Inside these vortices, the electric condensate vanishes and due to the Meissner effect produced by the vector boson mass, magnetic fields can only penetrate the superconductor medium through the interior of these vortices. Physically, these magnetic vortices must be closed because we are considering that the superconductor described by the Proca theory extends over the whole spacetime $\mathbb{R}^{1,3}$ and, according to the magnetic Gauss law of the Proca equations of motion \cite{joshi}, all the magnetic flux lines must be closed in the absence of monopoles. This picture, however, changes when we consider the presence of external probe monopoles, since open magnetic vortices are also formed in this case, having a monopole-antimonopole pair in their ends. To see this, we must consider what happens in regions with $\chi_2\neq 0$ (in regions where $\chi_2=0$ and $\Omega_2=\delta\lambda_3\neq 0$, we have from (\ref{eq:7}) the closed magnetic vortices disconnected from the monopoles, as just discussed). As stated before, the local magnetic Dirac brane symmetry corresponds to the freedom of deforming the magnetic Dirac branes through the space not occupied by the electric world-lines. In the electric condensed regime these world-lines proliferated such that they established a continuum corresponding to the electric condensate and thus, due to the Dirac's veto, the only place allowed for the magnetic Dirac strings is inside closed magnetic vortices formally connected to the monopoles. In such a setup, the flux inside the magnetic Dirac strings cancels out part of the flux inside the closed vortices, leaving as the result open magnetic vortices with a monopole-antimonopole pair in their ends. This is illustrated in figure \ref{fig:1}.

\renewcommand{\figurename}{Figure}

These open vortices are the physical confining magnetic flux tubes that emerged naturally in the formalism presented here due to a careful account of the local Dirac brane symmetry. Notice that via GJTA these confining flux tubes are consistently described by the Dirac brane invariants $L_2$ defined in (\ref{eq:7}), instead of the unphysical Dirac strings $\chi_2$. With this, the Dirac strings never become observable in our approach, contrary to what happens in other methods used in the literature \cite{qt,mvf,joshi,ahrens,ripka,antonov}.

We emphasize that the exact point that allows us to obtain a consistent formulation of monopoles in the Proca theory (and also in its electromagnetic dual, the massive Kalb-Ramond theory (\ref{eq:8})), without inconsistencies like ``observable Dirac strings'', is the consideration of the role played by the density of magnetic vortices, $\Omega_2=\delta\lambda_3$, which is present in the expression of the Dirac brane invariants $L_2$ defined in (\ref{eq:7}), but is missing in the vast majority of the literature on the subject. If one forgets about the magnetic vortex density $\Omega_2$, then $L_2$ reduces to the Dirac string term $\chi_2$, which is them placed directly over the electric condensate, and this is exactly the source of all the problems mentioned above: it violates the Dirac's veto, and therefore, violates the local Dirac brane symmetry. Indeed, the ensemble of magnetic vortices $\left\{*\lambda_3\right\}$ is physically essential: in the presence of external monopoles it is impossible to realize a complete electric condensation. Although the closed magnetic vortices disconnected from the monopoles can be completely diluted, the closed vortices connected to the monopoles cannot be undone: in fact, although the magnetic fields generated by the monopoles are expelled by the Meissner effect from almost the whole space constituted by the electric condensate, these fields cannot simply vanish and they become confined inside regions where the electric condensate has not been established, which correspond to open magnetic vortices with a monopole-antimonopole pair in their ends. These confining flux tubes are described by the Dirac brane invariants $L_2$, and not by the Dirac strings $\chi_2$, which are indeed removed from the formalism in the condensed phase, as illustrated in figure \ref{fig:1}.

Finally, let us obtain the effective interaction potential for the external monopoles. For this sake, we must first elaborate further about the formal sum over vortices present in the partition function (\ref{eq:10}). As discussed above, in the presence of external probe monopoles, there are in general both, closed vortices disconnected and closed vortices formally connected to the monopoles, with the latter giving rise to open vortices with monopole-antimonopole pairs in their ends through the Dirac string cancellation scheme explained in figure \ref{fig:1}. Therefore, we write in general: $\sum_{\left\{*\lambda_3\right\}}= \sum_{\left\{*\lambda_3^{\textrm{disc}}\right\}}\sum_{\left\{*\lambda_3^{\textrm{con}}\right\}}$, where $\sum_{\left\{*\lambda_3^{\textrm{disc}}\right\}}$ represents the formal sum over vortices disconnected from the monopoles and $\sum_{\left\{*\lambda_3^{\textrm{con}}\right\}}$ represents the formal sum over vortices connected to the monopoles. Notice that we must sum over all the possible configurations of vortices connected to the monopoles in order to assure that $L_2$ is a legitimate Dirac brane invariant variable. However, the ensemble of disconnected vortices is a somehow externally prescribed ensemble that should depend ultimately on some external control parameters like temperature, external magnetic fields, etc. If there are many or just a few of these vortices in the system and how is their dynamics, this is something that should be encoded in such a sum. For example, as it is well known, a vortex proliferation destroying a superconducting system (electric condensate) can be driven by either heating the system or by increasing an externally applied magnetic field. On the other hand, at low temperatures and weak external magnetic fields, there should be only a few (or none) vortices in the system. In this sense, the ensemble of disconnected vortices is externally prescribed. Via GJTA, we never approach the dynamics responsible for the proliferation or the dilution of topological currents (including vortices). We just assume that the system is in a given regime (diluted, condensed, or even something between these extremes) and look to construct effective field theories describing the lowest-lying modes for the regime under consideration, taking the associated symmetries as our main guide.

We do not know, in general, how to effectively evaluate the sum over the ensemble of disconnected vortices (although, at least one exception exists, as discussed in \cite{jt-schwinger}). Therefore, we shall consider in what follows a particular state of the system where the vortices disconnected from the monopoles are completely diluted. In this case, there only remains in the system a sum over closed vortices formally connected to the monopoles. As explained around figure \ref{fig:1}, this gives rise to the open vortices and, therefore, it is equivalent to sum over all possible shapes of world-surfaces of confining flux tubes, such that, in this case, the partition function (\ref{eq:10}) is reduced to:
\begin{align}
Z_c[j_1]&=\sum_{\left\{*L_2\right\}}^{\textrm{(open)}}\int\mathcal{D}A_1 \exp\left\{i\int_{\mathbb{R}^{1,3}}\left[-\frac{1}{2}(dA_1-g*L_2)\right.\right.\nonumber\\
&\left.\left.\wedge*(dA_1-g*L_2)+\frac{m^2}{2}A_1\wedge *A_1\right]\right\},
\label{eq:11}
\end{align}
where the sum $\sum_{\left\{*L_2\right\}}^{\textrm{(open)}}$ is taken over all the possible shapes of the world-surfaces of the confining magnetic flux tubes \cite{artigao}. Integrating out the massive vector field $A_1$ in (\ref{eq:11}), we obtain:
\begin{align}
Z_c[j_1]&=\exp\left\{i\int_{\mathbb{R}^{1,3}}-\frac{g^2}{2}j_1\wedge \frac{1}{-\Delta+m^2}*j_1\right\}\nonumber\\
&\sum_{\left\{*L_2\right\}}^{\textrm{(open)}}\exp\left\{i\int_{\mathbb{R}^{1,3}}\frac{m^2g^2}{2}L_2\wedge\frac{1}{-\Delta+m^2}
*L_2\right\},
\label{eq:12}
\end{align}
where we used that $j_1=\delta\chi_2=\delta L_2$. Considering a static external monopole-antimonopole configuration and the asymptotic time interval where the monopole-antimonopole pair is created in $t\rightarrow-\infty$ and is annihilated in $t\rightarrow+\infty$, the dominant contribution in the sum over configurations of the magnetic flux tubes in (\ref{eq:12}) is given by a straight tube \cite{artigao}, which gives the stable configuration of minimal energy of the system. In this limit we can take only this contribution into account such that the second term in (\ref{eq:12}) gives a magnetic confining potential that is linear in the monopole-antimonopole separation, while the first term gives a short-range Yukawa interaction. The static effective interaction potential between the external monopoles in this limit explicitly reads \cite{suganuma,chernodub-1} (see also \cite{artigao,ripka}):
\begin{align}
V_{eff}(R)=-\frac{g^2}{4\pi}\frac{e^{-mR}}{R}+\frac{m^2g^2}{8\pi}\ln\left(\frac{m^2+M^2}{m^2}\right)R,
\label{eq:13}
\end{align}
where $R$ is the distance between the monopole and the antimonopole and $M$ is a physical ultraviolet cutoff corresponding to the inverse of the coherence length of the electric condensate, which gives the thickness of the magnetic flux tubes. Hence, due to the magnetic confinement generated by the electric condensate, the introduction of isolated monopoles is impossible in the Proca theory, since this would render the energy of the system infinite (what can be seen from (\ref{eq:13}) by taking the limit $R\rightarrow\infty$), the only possibility being the introduction of monopole-antimonopole pairs connected by Dirac brane invariants corresponding to the physical confining magnetic flux tubes. Besides, notice that the condensate interpretation is essential here, since the physical cutoff $M$, which avoids an ultraviolet divergence in the effective potential (\ref{eq:13}), is a natural finite scale of the system corresponding to the inverse of the coherence length of the electric condensate, which may be small but is certainly non-vanishing. Notice also that in the limit $m\rightarrow 0$, the electric condensate is removed and the confining term of the effective potential (\ref{eq:13}) goes to zero, while the Yukawa potential reduces to the usual Coulomb potential of the massless electrodynamics.

%It is also worth mentioning the fact that, in spite of the presence of an electric condensate, the vacuum remains electrically neutral \cite{chernodub-2}: the presence of an electric condensate automatically implies in the appearance of an electric condensate with a charge of the same magnitude, but with opposite sign, such that the vacuum energy is lowered and its net electric charge is zero.

\section{Concluding remarks}
\label{sec:conclusion}

In this Letter we discussed how to introduce in a consistent manner Dirac monopoles into the Proca theory in $(3+1)$-dimensions. This model of massive photons can be induced due to a condensation of electric charges, and external magnetic monopoles are found to be confined in monopole-antimonopole pairs connected not by Dirac strings, but by physical Dirac brane invariants corresponding to open magnetic vortices. Actually, this can be seen as a general conclusion involving theories with massive vector bosons and Dirac monopoles, from which the Proca theory is a particular example. Another example vastly discussed in the literature \cite{ht,pisarski,diamantini,mcsmon,artigao} is the Maxwell-Chern-Simons theory in $(2+1)$-dimensions in the presence of magnetic instantons. Also in this system, the (topological) mass of the vector boson can be interpreted as emerging due to an electric condensate that breaks the discrete $P$ and $T$ symmetries \cite{santiago}, and the magnetic instantons are found to be confined by physical magnetic flux tubes described by Dirac brane invariants \cite{mcsmon,artigao}. Recently, we reached the same conclusion in the Maxwell-BF theory in $(3+1)$-dimensions \cite{mbf-m}, what tells us that Dirac monopoles and massive vector bosons are not incompatible at all, but instead, the mass of the vector bosons can be in general interpreted as arising due to an electric condensate, which confines external magnetic defects through the Meissner effect. Therefore, what is really inconsistent is the introduction of \emph{isolated} Dirac monopoles in theories with massive vector bosons. The monopole confinement obtained in these theories does not imply any kind of incompatibility between monopoles and massive vector bosons as well as the fact that quarks not appearing as asymptotic states in QCD, being confined in the interior of hadrons, does not represent any kind of incompatibility between quarks and QCD.

We close our comments by stressing the important fact that the issue of the observability of the brane invariants that confine the probe magnetic monopoles in the Proca theory has nothing to do with the presence or the absence of gauge invariance, as one can see from \cite{helayel}, where problems similar to those reported in \cite{joshi} were also observed, in spite of the fact that in \cite{helayel} the Maxwell-BF theory was used to restore the gauge symmetry of the theory describing interacting massive photons and Dirac monopoles in $(3+1)$-dimensions. As explained here and also in \cite{mcsmon,artigao,mbf-m}, the key point involved in the consistent formulation of Dirac monopoles in a theory with massive vector bosons regards a careful treatment of the Dirac brane symmetry in the passage from the regime with diluted electric charges to the regime with condensed electric charges, where the mass of the vector bosons emerges. It is essential, for a consistent formulation of such systems, to take into account the contribution of the magnetic vortices describing regions where the electric condensate has not been established. These magnetic vortices are always present when Dirac monopoles are inserted into models with massive vector bosons and neglecting them renders the formalism inconsistent, due to a explicitly violation of the local Dirac brane symmetry. Moreover, it also seems necessary, in order to avoid an ultraviolet divergence in the effective interaction potential between probe magnetic monopoles in these systems, that the mass of the vector bosons arises due to an electric condensate, since with this interpretation a finite ultraviolet cutoff scale corresponding to the inverse of the coherence length of the condensate is naturally contemplated in the formalism.

\acknowledgments

We thank Conselho Nacional de Desenvolvimento Cient\'ifico e Tecnol\'ogico (CNPq) for financial support.

\end{document}